# Nanoscale lattice heterostructure in high Tc superconductors

Annette Bussmann-Holder[1], Jürgen Haase[2], Hugo Keller[3], Reinhard K. Kremer[1], Sergei I. Mukhin[4,5], Alexey Menushenkov[6], Andrei Ivanov[6], Alexey Kuznetsov[6], Victor Velasco[7], Steven D. Conradson[8], Gaetano Campi[9,10], Antonio Bianconi[9,10]

[1] Max-Planck-Institute for Solid State Research, Heisenbergstr. 1, D-70569 Stuttgart, Germany (a.bussmann-holder@fkf.mpg.de, 0000-0003-1380-6399), (rekre@fkf.mpg.de, 0000-0001-9062-2361)

[2] Faculty of Physics and Earth Science, University of Leipzig, Germany (juergen.haase@uni-leipzig.de, 0000-0002-0140-6964)

[3] Physik-Institut der Universität Zürich, Winterthurerstr. 190, CH-8075 Zürich, Switzerland (keller@physik.uzh.ch, 0000-0001-8229-250X)

[4] Instituut-Lorentz, Universiteit Leiden, P.O. Box 9506, 2300 RA Leiden, The Netherlands (i.m.sergei.m@gmail.com, 0000-0002-3051-6927)

[5] Theoretical Physics and Quantum Technologies Department, National University of Science and Technology "MISIS," Leninski avenue 4, 119049 Moscow, Russia

[6] National Research Nuclear University "MEPhI" Kashirshskoe shosse, 31, 115409 Moscow (https://orcid.org/0000-0002-4252-5142, email: apmenushenkov@mephi.ru), (https://orcid.org/0000-0002-7904-3833, email: andrej.ivanov@gmail.com), (http://orcid.org/0000-0002-9279-8296, email: avkuznetsov@mephi.ru)

[7] International School for Advanced Studies (SISSA), Via Bonomea 265, 34136 Trieste, Italy (https://orcid.org/0000-0003-3904-8602, vvelasco@sissa.it)

[8] Department of Chemistry, Washington State University, Pullman, WA 90164, USA (https://orcid.org/0000-0002-7499-9465, st3v3n.c0nrads0n@icloud.com)

[9] Institute of Crystallography, National Research Council, CNR, Via Salaria Km 29.3, 00015 Monterotondo Rome, Italy (https://orcid.org/0000-0001-9845-9394, gaetano.campi@cnr.it)

[10] Rome International Center for Materials Science Superstripes RICMASS, Via dei Sabelli 119A, 00185 Rome, Italy (https://orcid.org/0000-0001-9795-3913, antonio.bianconi@ricmass.eu)

\* *Corresponding authors:* a.bussmann-holder@fkf.mpg.de; antonio.bianconi@ricmass.eu, juergen.haase@uni-leipzig.de, gaetano.campi@cnr.it, i.m.sergei.m@gmail.com, apmenushenkov@mephi.ru, vvelasco@sissa.it, st3v3n.c0nrads0n@icloud.com

**Abstract** Low temperature superconductivity was known since 1957 to be described by BCS theory for an effective single band metals controlled by the density of states at the Fermi level, very far from band edges, the electron phonon coupling $\lambda$, and the energy of the boson in the pairing interaction $\omega_0$, but BCS has failed to predict high temperature superconductivity in different materials above about 23 K. High temperature superconductivity above 35 K since 1986 has been a matter of materials science where manipulating the lattice complexity of high temperature superconducting ceramic oxides (HTSC) has driven material scientists to grow new HTSC quantum materials up to 138K in $HgBa_2Ca_2Cu_3O_8$ (Hg1223) at ambient pressure and near room temperature in pressurized hydrides. This perspective covers the major results of materials scientist in these last 39 years investigating the role of lattice inhomogeneity detected in these new quantum complex materials. We highlight the nanoscale heterogeneity in these complex materials and elucidate their special role played in the physics for HTSC. Especially, it is pointed out that the geometry of lattice and charge complex heterogeneity at nanoscale is essential and intrinsic in the mechanism of rising quantum coherence at high temperature

# 1. Essential Heterogeneities in Hole-Doped high Tc Cuprate Superconductors

Only a year after their revolutionary work in superconductivity, J. G. Bednorz and K. A. Müller were awarded the Nobel prize for their "important break-through in the discovery of superconductivity in ceramic materials" [1]. This wording implies already that the material, namely "ceramics", is an inhomogeneous off-stoichiometric compound. A clear proof of inhomogeneity was rapidly provided by site-selective oxygen isotope experiments on the superconducting transition temperature $T_C$ and on phonon frequencies, emphasizing that structurally different oxygen ions in YBaCuO (chain, apex and plane oxygens) contribute differently to the total oxygen isotope effect [2, 3]. In Zürich oxygen isotope effects were further investigated and a number of unexpected effects on the superconducting properties were discovered [4 – 10]. The differentiation between constituting structural elements of the cuprates continued with the seminal work by the Bianconi group on of $La_{1.85}Sr_{0.15}CuO_4$ [11,12] where stripe like patterns referring to tetragonal and orthorhombic structural elements have been detected. This work prompted intense efforts in the search for heterogeneity and phase separation [13 – 21] in cuprates and initiated a novel understanding by showing that at least two superconducting order parameters, namely s + d, coexist [22, 23]. Experimentally as well as theoretically a broad number of papers were devoted to this subject and its existence was convincingly demonstrated.

K. A. Müller and J. G. Bednorz started their project from the idea that polarons in perovskites and the Jahn-Teller effect are at the origin of high temperature superconductivity [24] (see figure 1). Local lattice effects were in the focus in the following years and especially the work of the Bianconi group [11, 12] was continued and a coexistence of localized carriers with itinerant ones could be established. Importantly it could be shown that the onset temperature T* of stripe formation carries a huge oxygen isotope effect [5, 10, 25] which again was dependent on the structurally different elements. Coinciding with this finding were local anomalous lattice effects evident from phonon anomalies and neutron scattering [26]. These results prove that heterogeneity and local lattice effects have been in the focus from 1986 on and proposed and experimentally validated since almost 40 years.

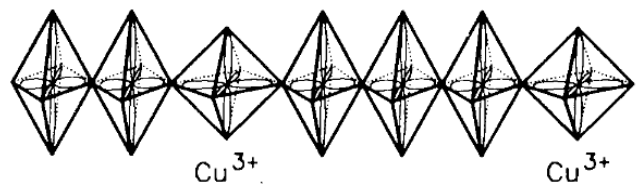

**Figure 1.** Schematic sketch of the Jahn-Teller polaron (from Ref. 1).

## 2. Q-balls and the role of local nanoscale lattice fluctuations in the high Tc superconductivity mechanism in cuprate perovskites

The recently proposed mechanism of the pseudogap state and high-Tc superconductivity in cuprates [27-29] was supported by data obtained in $HgBa_2CuO_{4+y}$ via micro X-ray diffraction using synchrotron radiation techniques [30-31]. The quintessence of the Q-balls (nontopological solitons) mechanism proposed for localized electronic states in cuprates lies in the idea that nested fermionic states on the Fermi surface of the electron/hole subsystem may cause an instability in the form of local condensation of boson-ic spin- or charge fluctuations, that break symmetry of the Matsubara time chirality in the Euclidean space-time forming the Q-balls (nontopological solitons) predicted by Sidney Coleman [32] for quark-gluon plasma in Minkowski space-time. These local collective bosonic fluctuations, arising via the first order phase transition below the temperature T* [27-28], possess wave-vectors that connect the nested fermionic states at the opposite 'antinodal' regions of the Fermi surface forming superconducting condensates of Cooper /local pairs inside the Q-balls, thus lowering their energy via opening of the energy gap on the nested parts of the Fermi surface. Hence, the pseudogap phase is created. The amplitude of the bosonic field inside the Q-ball rotates either clockwise or anticlockwise with Matsubara frequency $\Omega=2\pi T$ in the Euclidean space-time [27-29]. Finite radius solution for the Q-ball nontopological soliton is allowed due to conservation of the Noether 'charge' Q proportional to the number of condensed spin- or charge excitations forming the Q-ball [27,32]. It was found recently [33] that linear temperature dependence of electrical resistivity arises naturally due to scattering of not condensed fermions on the local Q-ball spin- charge fluctuations, that may explain famous 'Planckian' behavior of the 'strange metal' phase in high-Tc cuprates observed experimentally [34]. The diamagnetic response of Q-balls gas is also calculated [33] and shows good accord with experimental data by L. Li et al. [35] in the 'strange metal' phase. The phase diagram of high-$T_c$ cuprates with superconducting dome touching with 'strange metal' area at the optimal holes doping is also reproduced by the theory within Q-ball mechanism [27-29,33]. In total, we believe, these results provide support to the quantum thermodynamic time crystals, the Euclidean Q-balls, as the model of nanoscale lattice heterostructure driving superconducting properties of high-Tc cuprates.

## 3. Local lattice fluctuations in cuprates seen by X-ray spectroscopy XANES and EXAFS

X-ray absorption spectroscopy (XANES and EXAFS) at synchrotron radiation has proven to be the most informative for studying the local structure of high-temperature superconductors (HTS), which was first demonstrated on the $BaBiO_3$-based superconducting oxide family two years before the discovery of HTS in cuprates [36]. Polarized XANES and EXAFS studies at the K-Cu edge of thin copper-based HTS films, irradiated with high-energy helium ions, allowed for the first time to obtain data on changes in the local electronic structure in the Cu-$O_2$ plane, leading to the loss of superconducting properties [37—41]. A unique experiment on the study of the electronic structure of $Nd_{1.85}Ce_{0.15}CuO_{4-\delta}$ films irradiated by $He^+$ ions using XANES at Cu-$L_3$ and Ce-$M_{4,5}$ edges, was performed at the superconducting synchrotron Super-ACO (LURE, Orsay, France). It provided new information concerning the role of oxygen deficiency in the emergence of superconducting properties of electron-doped cuprates [42,43]. The low-temperature studies of the EXAFS spectra in $La_{2-x}Sr_xCuO_4$ and $Nd_{2-x}Ce_xCuO_{4-\delta}$ revealed low-temperature anharmonicity manifested as anomalous temperature dependences of the Debye-Waller factors of the Cu-O bond in the superconducting Cu-$O_2$ plane. The observed anomalies were explained by the oxygen ion vibrations in a double-well potential at low temperatures for both hole- and electron-doped cuprates [44—46], similar to those previously observed in the family of high-temperature superconductors based on $BaBiO_3$ [47]. The results of low-temperature studies allowed to propose a model of the mechanism of superconductivity in cuprates based on local charge carrier pairing in real space [48] by analogy with the bismuthate family [47]. The achieved understanding of the features of the local structure of cuprates made it possible to determine the correlations between the influence of nano-inclusions that increase the critical current in real 2G MOCVD YBCO tapes and the changes these inclusions introduce into the local structure of the Cu-$O_2$ plane of HTS-coated conductors [49,50].

In addition it has been shown that the local non-centrosymmetric structure of $Bi_2Sr_2CaCu_2O_{8+y}$ by X-ray magnetic circular dichroism at Cu $K$-edge XANES [51].

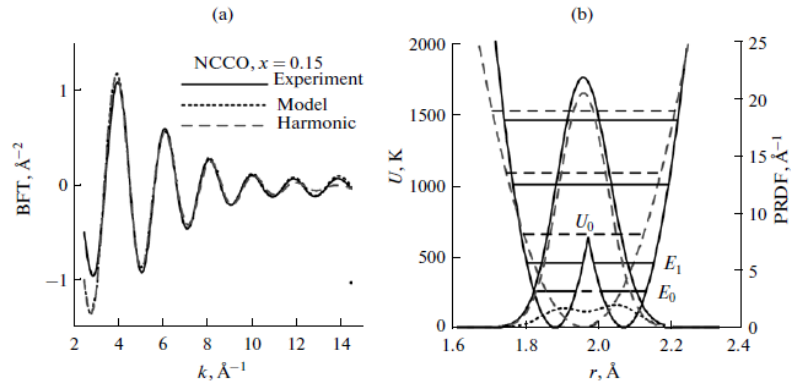

**Figure 2**. Panel (a) EXAFS oscillations above the Cu $K$ edge in $Nd_{1.85}Ce_{0.15}CuO_{4-\delta}$ and panel (b) the extracted double-well potential for oxygen ion vibration (from Ref. 45).

## 4. Internal Quantum Tunneling Polarons Dynamical structure and Kuramoto synchronization in cuprate superconductors

The microscopic mechanism behind high-temperature superconductivity (HTSC) in cuprate superconductors remains elusive, despite decades of experimental and theoretical efforts. However, there is a growing body of evidence that suggests that charge-lattice dynamics associated with the apical oxygen ($O_{ap}$) atoms play an important role. In this context, recent studies combining Extended X-ray Absorption Fine Structure (EXAFS) spectroscopy and exact diagonalization modeling have provided a novel framework for understanding the coupling between local charge inhomogeneities and anharmonic lattice dynamics in terms of Internal Quantum Tunneling Polarons (IQTPs) [52-54]. IQTPs differ from conventional, e.g., Holstein-Hubbard polarons, in that the excess charge is localized on the oxygen instead of the metal ion. This adds charge transfer between neighboring oxygen ions, accompanied by the associated shift in the bond length, to the polaron dynamics. The transit through the lattice is therefore quantum tunneling rather than hopping. Recent EXAFS experiments on heavily overdoped cuprates such as $YSr_2Cu_{2.75}Mo_{0.25}O_{7.54}$ and $Sr_2CuO_{3.3}$ [55,56] have extended the original ones [57-65] to reveal temperature-dependent transformations in the local structure at the superconducting (SC) transition not only in the Cu-$O_{ap}$ distances and numbers, but also in the Cu-Sr anharmonicity. These features are captured in real space via deviations from Gaussian pair distributions in the Fourier-transformed spectra signaling the presence of soft, anharmonic vibrational modes [57-60]. These distortions are not associated with thermal effects, instead they emerge from the tunneling of the charge and lattice distortion between degenerate Cu-O configurations. The defining characteristic of the IQTP is therefore that the charge and atomic displacement oscillate between sites faster than the polaron's center-of-mass hopping rate, confining the dynamics within a small cluster. The defining signature of an IQTP that differentiates it from a conventional polaron is its presence in the inelastic structure, S(Q, ), at the energy corresponding to the tunneling frequency. It is therefore observed in EXAFS that it originates in the instantaneous structure, S(Q,t=0), but not in the Bragg peaks of the diffraction pattern that are determined by S(Q, =0), that is, the elastic structure. Facilitated by its element specificity and magnetic orientation of samples, IQTPs are therefore identified in EXAFS as a two-site distribution of a Cu-$O_{ap}$ pair that is not observed in diffraction or elastic pair-distribution functions. These features have been observed over the years in various hole-doped cuprates [61-71]. However, because of the limited number of scientists familiar with inelas-tic structure and EXAFS as a probe of it, it has often been attributed to static disorder or local lattice inhomogeneities to the exclusion of the importance of dynamic structure and IQTPs. Notwithstanding this opinion by the larger community, we have proceeded with interpreting the multi-site Cu-O distributions that are coupled to the SC transition as a signature of collective quantum coherence. Specifically, the six-atom cluster model that comprises two Cu-Oap IQTPs coupled via a shared planar oxygen (Opl), and a soft Oap-Sr-Oap bridge, exhibits a first-order phase transition triggered by an anharmonic three-phonon interaction mediated by the Sr ion, coupling the two IQTPs, to a phase-synchronized state [52-54], like the one described by the Kuramoto model in networks of coupled oscillators [67-69]. This synchronization causes an expansion of the charge distribution from its location on the apical sites into the $CuO_2$ plane. This increase of the planar electronic density may be a requirement for pair formation. At the same time, a planar IQTP mode develops in the planar oxygen site, suggesting that coherent charge-lattice dynamics are not restricted to the dielectric layer but extends into the superconducting planes. These findings were further confirmed by mapping the full quantum Hamiltonian of the cluster onto a Kuramoto-like mean-field equation, revealing an emergent order parameter that quantifies the internal synchronization of polarons [52-54].

These results put into question the so-called "passive" role of the dielectric layer in cuprates. The traditional view of the Sr-$O_{ap}$ layer as limited to a structural spacer or charge reservoir is challenged by evidence showing that anharmonic dynamics and charge redistribution within this layer are strongly correlated with superconducting properties [70,71]. The IQTP unifies diverse observations from EXAFS, Raman, and inelastic neutron scattering, and it offers a microscopic picture of nonadiabatic, strongly coupled lattice-charge dynamics. Moreover, the synchronization of IQTP's offers a route to understanding how local lattice dynamics involving apical oxygen can participate in the high-Tc mechanism. An essential next step is to determine whether the coherence inherent to the entire six-atom system extends further, where it may promote the formation of the coherent superfluid. By bridging real-space structural probes with quantum synchronization theory, this approach opens new avenues for interpreting complex behavior in quantum complex matter and motivates the design of new compounds and techniques where local dynamical coherence may be tuned.

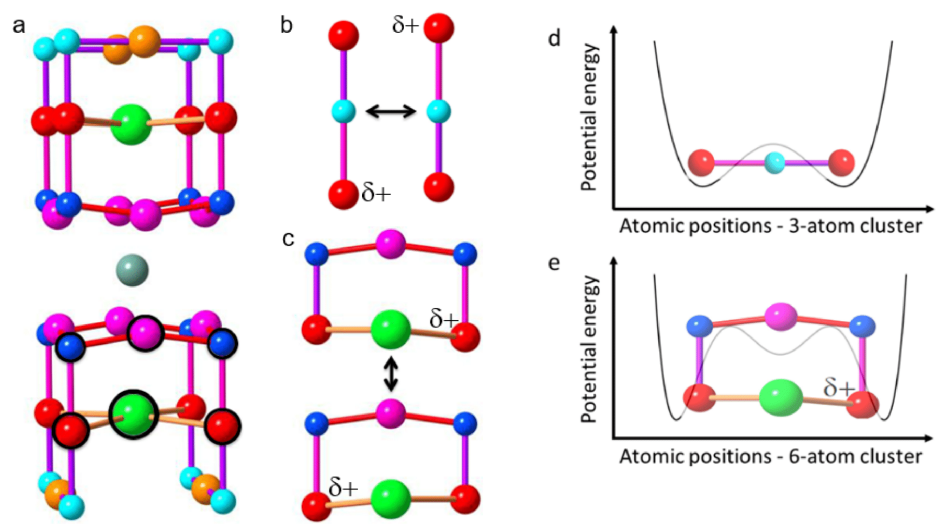

**Figure 3**. a) The crystal structure with the general features of multilayered cuprates with higher Tc's. b) The original three-atom cluster derived from the two-site distributions [64]. c) The atoms circled in black in (a) form the six-atom cluster [52-54]. d) For the three-atom cluster the potential energy corresponds to a double-well structure [64]. e) For the 6-atom cluster, the potential energy corresponds to a triple-well structure.

## 5. Polarons in cuprate perovkites as seen by X-ray Absorption Spectroscopy

The polaronic states are related with the breakdown in quantum complex materials of the Born-Oppenheimer approximation used in standard solid state physics to separate the motions of free electrons and nuclei. In 1986 Alex Müller and Georg Bednorz proposed that high-temperature superconductivity appears in the complex phase of doped cuprate perovskite oxides by condensation of Jahn Teller bipolarons [72-75] expected near Metal-insulator transitions in strongly correlated electronic systems [5] which have been observed recently [77]. The $CuO_2$ plane in doped cuprates was described as a mixed valence phase of a transition metal oxide made of a background of $Cu^{2+}$ ($3d^9$) states and Jahn Teller polaron associated with hole doping. To underpin the complex phase behavior of polarons controlled by the interplay between charge carriers and lattice distortions. X-ray absorption spectroscopy techniques, particularly extended X-ray absorption fine structure (EXAFS) and X-ray absorption near-edge structure (XANES), have emerged in the eighties of the XX century as new powerful tools to directly probe the local lattice environment around copper sites in these materials. [11, 78-82].

The microscopic lattice evidence for self organization of pseudo Jahn-Teller polarons [72, 83] associated with the hole hoping formed by $Cu(3d^9)O(2p^5)$ singlet configuration called $3d^9\underline{L}$ (where $\underline{L}$ indicate a ligand hole in the oxygen plaquette) formed by mixing of Cu $3d_{x^2-y^2}$ and $3d_{z^2}$ orbitals with $b_1$ and $a_1$ orbitals of the oxygen plaquette with the polaronic pseudo-Jahn Teller local lattice distortions (LLD) organized at nanoscale was provided by a very fast and local experimental method X-ray absorption spectroscopy [83]. Central to the polarons detection of two distinct Cu–O planar bond lengths within the $CuO_2$ planes of $La_{1.85}Sr_{0.15}CuO_4$ was revealed by polarized Cu K-edge EXAFS at low temperatures below approximately 100 K. This splitting signals the coexistence of two types of $CuO_6$ octahedra: the first one with a small tilt of rectangular $CuO_4$ planar plaquette (LTO-like) with a long Cu-$O_{ap}$ distance the second one with a rhombic $CuO_4$ planar plaquette with a larger tilt angle (LTT-like) and short Cu-$O_{ap}$ distance, corresponding to regions where doped holes locally trap lattice distortions, thus forming D stripes of condensed pseudo-Jahn Teller polarons (see Figure 4). The probability of (LTT-like) domains appear to be consistent with a substantial fraction (~33%) of the rhombic $CuO_4$ plaquettes developing local distortions in D stripes and (~66%) LTO-like plaquettes forming U stripes. Such coexistence creates a superlattice of striped phases called "superstripes" scenario: nanoscale arrays of distorted D and undistorted U lattice stripes. The Superstripes scenario is thought to be a real-space manifestation of polaronic charge ordering reported quite independently by Goodenough [84,85] and predicted by Kusmartsev et al. theoretical analysis [86]. These local lattice distortions are connected with the emergence of a striped phase, evidenced by complementary diffuse X-ray scattering measurements that reveal superlattice modulations indicative of nanoscale periodic lattice distortions. The Bianconi-Perali-Valletta (BPV) theory of high Tc superconductivity [87] was developed on the basis of the phenomenological data on the nanoscale geometrical parameters of the observed topology of the stripes phase shown in Fig.2. This theory has predicted the Tc amplification by the Fano-Feshbach shape resonance in multi-gap superconductivity where nanoscale superconducting units are quantum confined by intercalated normal units disclosed in the patent [88] opening the road map to quantum material design of artificial high Tc superconductors [89-93].

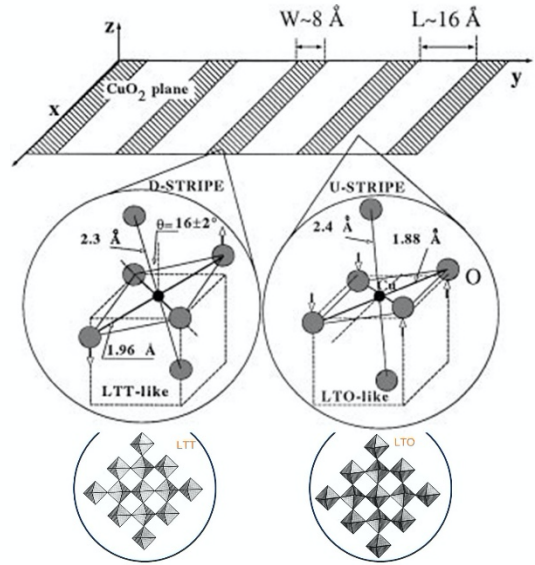

**Figure 4**. Pictorial view of the nanoscale phase separation of stripes made of first units, polaronic doped distorted $CuO_6$ octahedra, left side, with the "LTT like" structure assigned to the doped distorted D-stripes) of width W and second units of the undistorted octahedra, right side, of the undoped "LTO like" undistorted stripes (U stripes) of width L. The period of the superlattice of quantum stripes is d=L+W=2.4 nm and the quantum geometry factor L/d at optimum doping show the magic value 2/3=0.66.

It has been shown that nanoscale phase separation could be associated with the proximity of the Fermi level to a Lifshitz topological transition for the appearance of new Fermi surface in a correlated multi-band system [94]. The polaronic stripes in cuprates reflect a charge density wave intertwined with an orbital and bond order wave, a complex electronic-lattice ordering stabilized by strong anisotropic electron-phonon coupling as detailed in [76] and [95] which emphasizes that the polaronic CDW phase in cuprates can be associated by a generalized Van der Waals phase separation thermodynamic model where doped holes behave as interacting extended objects exhibiting directional bonding tendencies around a critical hole doping level of 1/8. Below this doping, the system forms localized Mott insulating puddles coexisting with striped polaronic CDW puddles, while above this doping range, striped CDW puddles coexist with metallic (Fermi liquid) puddles, yielding a nanoscale phase separation landscape. The model accounts for these phase coexistences as a gas-to-liquid-like transition in a polaronic charge fluid and is supported by X-ray absorption and diffraction data demonstrating the intricate pattern of puddles and stripes in perovskites. Furthermore, experimental findings show that the CDW onset temperature with its lattice character can possess a giant isotope effect [96,97], strongly implicating lattice vibrations in the stabilization of polaronic CDW phases. Additionally, the misfit strain between the active copper oxide layers and spacer layers strongly influences the local lattice distortions and promotes an arrested phase transition manifesting as a correlated quantum disorder phase of stripe puddles in perovskites.

Crucially, EXAFS and XANES are sensitive probes, detecting the amplitude of the periodic lattice distortion and its spatially heterogeneous distribution, thereby providing direct evidence for intrinsic nanoscale electronic and lattice inhomogeneity which has been unveiled in the superconducting overdoped cuprate [98-102].

The resulting "superstripes" phase exhibits a complex multiscale patterning across length scales ranging from the atomic to mesoscale, reflecting a frustrated electronic and lattice phase separation where polaronic and metallic regions coexist. This arrested phase separation scenario helps explain the broad transition widths and the competition between superconductivity and charge ordering.

Collectively, the combination of EXAFS/XANES results and complementary diffraction studies consolidates the viewpoint that polarons in cuprates manifest as lattice-localized distortions that organize into nanoscale striped CDW puddles, intimately linked to the physics of the Mott insulator-to-metal crossover and high-temperature superconductivity itself. The detailed temperature- and doping-dependent evolution of these polaronic distortions documented in the referenced studies provides a consistent microscopic foundation for understanding the complexity of cuprate phase diagrams, emphasizing the essential role of lattice degrees of freedom concomitant with strong electron correlations. Recent advances have shown a roadmap to high temperature superconductivity by tailoring nanoscale lattice heterostructures

The comprehensive spectroscopic evidence underscores that superconductivity in cuprates emerges from a complex nanoscale electronic-lattice landscape, where polaronic charge density waves and spatial phase separation are fundamental elements as revealed uniquely by X-ray absorption spectroscopy techniques.

## 6. Conclusions

Finally X-ray absorption and advanced diffraction methods have been able to unveil polaron formation and complex landscape of static polaron self organization in manganites, [103.105] and nickelates [106,107] in cuprates these fast and local methods have clearly shown the presence of coexisting local polaronic states and itinerant states in a complex phase of quantum matter showing nanoscale phase separation with the emergence of high temperature superconductivity [12,17,29,33,90,93].

**Author Contributions:** *Chapter 1* was written by Annette Bussmann-Holder, Jürgen Haase, Hugo Keller, and Reinhard K. Kremer. *Chapter 2* was written by Sergei I. Mukhin. *Chapter 3* was written by Alexey Menushenkov, Andrei Ivanov, Alexey Kuznetsov. *Chapter 4* was written by Victor Velasco and Steven D. Conradson, *Chapter 5* was written by Gaetano Campi and Antonio Bianconi.